\documentclass[final,5p
,twocolumn,authoryear]{elsarticle}

\usepackage{amsmath,amsfonts,amssymb,mathtools}
\usepackage{hyperref}% add hypertext capabilities
\usepackage[nameinlink,capitalise]{cleveref}
\usepackage{bm}
\usepackage{xcolor}

\usepackage[scr]{rsfso}
\newcommand{\pd}{\partial}
\newcommand{\de}{\mathrm{d}}
\newcommand{\cH}{\mathcal{H}}
\newcommand{\cZ}{\mathcal{Z}}
\renewcommand{\pd}{\partial}
\newcommand{\ii}{\mathrm{i}}
\newcommand{\bmag}{\mathcal{Q}}
\newcommand{\bevo}{\mathcal{E}}

\newcommand{\physrep}{Physics Reports}
\newcommand{\jcap}{J. Cosmol. Astrop. Phys.}
\newcommand{\mnras}{Mon. Not. Roy. Astron. Soc.}
\newcommand{\aap}{Astron. Astrophys.}
\newcommand{\prd}{Physical Review D}
\newcommand{\prl}{Physical Review Letters}
\newcommand{\aj}{AJ}
\newcommand{\apj}{The Astrophysical Journal}
\newcommand{\apjl}{The Astrophysical Journal Letters}
\newcommand{\nar}{New Astronomy Reviews}

\newcommand{\fluxum}{\times10^{-16} \rm \; erg\,cm^{-2}\,s^{-1}}

\usepackage{cancel}

\def\review#1{\textcolor{black}{#1}}
\def\reviews#1{\textcolor{black}{#1}}

\def\done{\textcolor{orange}{$\spadesuit$}} % FM

\journal{Physics of the Dark Universe}

\begin{document}

\begin{frontmatter}



\title{Detecting Relativistic Doppler by Multi-Tracing a Single Galaxy Population} %% Article title


\author[label1,label2]{Federico Montano}
\ead{federico.montano@unito.it}
\author[label1,label2,label3,label4]{Stefano Camera}
\ead{stefano.camera@unito.it}

\affiliation[label1]{organization={Dipartimento di Fisica, Università degli Studi di Torino},
             addressline={Via P.\ Giuria 1},
             city={Torino},
             postcode={10125},
             country={Italy}}
\affiliation[label2]{organization={INFN -- Istituto Nazionale di Fisica Nucleare, Sezione di Torino},
             addressline={Via P.\ Giuria 1},
             city={Torino},
             postcode={10125},
             country={Italy}}
\affiliation[label3]{organization={INAF -- Istituto Nazionale di Astrofisica, Osservatorio Astrofisico di Torino},
             addressline={Strada Osservatorio 20},
             city={Pino Torinese},
             postcode={10025},
             country={Italy}}
\affiliation[label4]{organization={Department of Physics \& Astronomy, University of the Western Cape},
             city={Cape Town},
             postcode={7535},
             country={South Africa}}

%% Abstract
\begin{abstract}
New data from ongoing galaxy surveys, such as the \textit{Euclid} satellite and the Dark Energy Spectroscopic Instrument (DESI), are expected to unveil physics on the largest scales of our universe. Dramatically affected by cosmic variance, these scales are of interest to large-scale structure studies as they exhibit relevant corrections due to general relativity (GR) in the $n$-point statistics of cosmological random fields. We focus on the relativistic, sample-dependent Doppler contribution to the observed clustering of galaxies, whose detection will further confirm the validity of GR in cosmological regimes. Sample- and scale-dependent, the Doppler term is more likely to be detected via cross-correlation measurements, where it acts as an imaginary correction to the power spectrum of fluctuations in galaxy number counts. We present a method allowing us to exploit multi-tracer benefits from a single data set, by subdividing a galaxy population into two sub-samples, according to galaxies' luminosity/magnitude. To overcome cosmic variance we rely on a multi-tracer approach, and to maximise the detectability of the relativistic Doppler contribution in the data, we optimise sample selection. As a result, we find the optimal split and forecast the relativistic Doppler detection significance for both a DESI-like Bright Galaxy Sample and a \textit{Euclid}-like H$\alpha$ galaxy population.
\end{abstract}



%% Keywords

\end{frontmatter}



%% Use \subsubsection, \paragraph, \subparagraph commands to 
%% start 3rd, 4th and 5th level sections.
%% Refer following link for more details.
%% https://en.wikibooks.org/wiki/LaTeX/Document_Structure#Sectioning_commands
\section{Introduction} \label{sec1intro}

The standard cosmological model ($\Lambda$CDM) pictures a universe dominated by the so-called dark components: dark matter and dark energy, which together account for about $95\%$ of the total energy contribution. Despite its success in reproducing many cosmological observations, it is challenged by several issues \citep{2022JHEAp..34...49A,2022NewAR..9501659P,2016Univ....2...23D}. Therefore, it has been posed that those problems, as well as the dark components themselves, may be due to an incorrect use of the theory to fit experimental data. General Relativity (GR) provides us with a description of the gravitational interaction which has passed various stunning tests \citep{2010JHA....41...41H,1968PhRvL..20.1265S,2002PhT....55e..41A,1975ApJ...195L..51H,2005ASPC..328...25W,2023ApJ...951L...8A,2016PhRvL.116f1102A}, and has been established as the theoretical framework depicting cosmic evolution---as though it were the game rules which the whole content of the universe has been following. However, GR is yet to be confirmed in cosmological regimes, characterised by huge distances and extremely weak gravitational fields. Since such conditions are common across the universe, the $\Lambda$CDM model in fact relies upon the assumption of GR correctly describing gravity even at those still unprobed scales. For this reason, there is a considerable interest in setting up tests that could either falsify or further confirm GR on cosmological scales. 

The large-scale structure (LSS) of the universe can be assessed as the product of cosmic evolution. In particular, LSS statistical properties, being both predictable and measurable, offer us the opportunity to look for GR-induced effects. We usually map biased tracers, e.g.\ galaxies, in redshift space to track the underlying (dark) matter distribution, so that a variety of phenomena come into play as corrections between the two distributions. Among those, we find the well-known clustering term \citep{2018PhR...733....1D}, the Redshift Space Distortions \citep{10.1093/mnras/227.1.1}, and a plethora of relativistic effects predicted by GR \citep{2010PhRvD..82h3508Y,2011PhRvD..84f3505B,2011PhRvD..84d3516C}, whose leading contribution is given by the relativistic Doppler term. A measurement of the latter could confirm the validity of GR on large scales, though it has not been achieved yet by the ongoing galaxy surveys. The Doppler contribution, being relevant on the very scales plagued by cosmic variance, calls for new observational campaigns \citep{2016arXiv161100036D,2013LRR....16....6A,2024arXiv240305398M,2022arXiv220904322S} and strategies to be detected. For the sake of completeness, other effects enter the computation of the observed spatial distribution of sources: mostly Primordial Non-Gaussianity \citep{2014PDU.....5...75M,2004PhR...402..103B} and Wide-angle effects \citep{2018MNRAS.476.4403C}. Nevertheless, we focus on the Doppler contribution in this paper.

To probe LSS properties we study the two-point statistic in Fourier space---i.e.\ the galaxy power spectrum---and address the relevance of the relativistic effects. The Doppler term can be investigated via a multi-tracer approach, that is, analysing the signal coming from different tracers at once. On the one hand, it is known that multi-tracer power spectra are able to boost the relativistic contribution thanks to the study of the cross-correlations, which show a milder scale dependence of the Doppler term \citep{2009JCAP...11..026M}. On the other hand, it allows us to even overcome the low statistical sampling limitation at large scales (cosmic variance), as was firstly noted in \citet{2004MNRAS.347..645P} \citep[see also][]{2009MNRAS.397.1348W,2009JCAP...10..007M,2009PhRvL.102b1302S,2013MNRAS.432..318A,2015ApJ...812L..22F,2015PhRvD..92f3525A}{}{}.

Interestingly, the GR contribution happens also to be dependent upon the specifications of the target sample, like the slope of the luminosity function. Since different galaxy populations have different power spectra, it is necessary to account for those specific aspects in order to estimate the presence of a Doppler signal in the data \citep{2021JCAP...12..009M}. A kind of optimisation work can hence be carried out, with the aim of searching for a promising target to look at. For instance, low-$z$ probes are expected to exhibit a significant Doppler effect, the lensing term being negligible.

Furthermore, it is possible to fulfil the multi-tracer requirements, namely, to consider at least two tracers simultaneously, with a single galaxy population if a luminosity cut is applied. Splitting a galaxy sample into a faint and a bright sub-sample, according to luminosity, enables us to make cross-correlation measurements within only one data set \citep{2014PhRvD..89h3535B,2016JCAP...08..021B,2017JCAP...01..032G,2021JCAP...12..003F,2023arXiv230604213B}. The using of a multi-tracer approach, either within a single galaxy population or combining different tracers, has already been proven as an effective strategy to improve measurement uncertainties, both with mock and real data \citep{2013MNRAS.436.3089B,2014MNRAS.437.1109R,2016MNRAS.455.3230B,2016MNRAS.455.4046M,2020MNRAS.498.3470W,2021MNRAS.504...33Z,2021MNRAS.503.1149Z,2022JCAP...04..021M,2024JCAP...01..008M}. Nonetheless, we focus on the relativistic signal in this analysis.

In fact, this paper comes out as a follow-up work of \citet{2023arXiv230912400M}, which was focused on exploiting the benefit of cross-correlation measurements against auto-correlation ones. Now, we explore the potential of the multi-tracer technique and provide, for the first time, a comprehensive theoretical estimation of all the relevant quantities to study the impact of the Doppler signal in a full power spectrum analysis. In doing so, we use the faint-bright division strategy and adopt an information (or Fisher) matrix approach.
We present forecasts regarding the probability of detecting the relativistic contribution within two samples: a DESI-like low redshift Bright Galaxy Sample (BGS), and a high-$z$ \textit{Euclid}-like H$\alpha$ emitter population \citep{2023AJ....165..253H,2024arXiv240513491E}.

This paper is structured as follows: in \cref{sec:definitions} we define the multi-tracer power spectrum and its covariance; in \cref{sec:data} we describe our data sets, outlining the features of the luminosity cut technique as well as the studied galaxy populations. Then, we explain the information analysis formalism used in \cref{sec:analysis} and show the results in \cref{sec:results}. Finally, we draw our conclusions in \cref{sec:conclusions}.

%----------------------------------------------------------------------------------------------%
\section{Multi-tracer power spectrum} \label{sec:definitions}
Galaxy surveys map the extragalactic sky in some `observed' space. In the traditional jargon of galaxy clustering, this is `redshift space', to emphasise that the reconstruction of galaxy positions along the line of sight is affected by spurious peculiar velocity contributions to their cosmological redshift, resulting into an anisotropy in the signal between the radial direction and that transverse to it, globally referred to as `redshift-space distortions' (RSDs). In fact, observed radial positions are not the only ones to differ from real galaxy locations. Most notably, observed transverse positions are affected by gravitational lensing distortions. To these two main perturbations, other subdominant ones are also present, arising in a full general relativistic treatment due to de-projection from observations on the past light-cone \citep{2011PhRvD..84f3505B,2010PhRvD..82h3508Y,2011PhRvD..84d3516C}.
Therefore, the number density contrast of galaxy counts is known to be affected by several terms. Many studies have been carried out about the derivation and physical meaning and impact of the various corrections \citep{2015CQGra..32d4001J,2015MNRAS.451L..80C,2016JCAP...01..016D,2020JCAP...09..058D,2020JCAP...07..048B,2022JCAP...01..061C,2022MNRAS.510.1964M,2023PhRvL.131k1201F}.

In this work, we focus on the leading local contributions to the number density. The density contrast of fluctuations in galaxy number counts then reads
\begin{equation}
\label{eq:delta_g_real_space}
    \varDelta(\bm x)=b\,\delta(\bm x)-\frac{1}{\cH}\,\pd_\parallel v_\parallel(\bm x)-\alpha\,v_\parallel(\bm x)\;,
\end{equation}
where $\delta$ is the matter density contrast (in comoving-synchronous gauge), $b$ is the (linear) galaxy bias, $\cH$ is the conformal Hubble factor, $\bm v$ is the peculiar velocity field, and the subscript `$\parallel$' denotes the component of a vector along the line of sight (which is always oriented from the observer towards the source). The first term in \cref{eq:delta_g_real_space} is a purely clustering term \citep{2018PhR...733....1D}, the second is the RSD term \citep{10.1093/mnras/227.1.1}, and the last one is the relativistic Doppler. The amplitude of the last term is
\begin{equation}
    \alpha\coloneqq\frac{\cH'}{\cH^2}+\frac2{r\,\cH}+2\,\bmag\left(1-\frac1{r\,\cH}\right)-\bevo\;,\label{eq:AD}
\end{equation}
with a prime denoting derivation with respect to conformal time, $r$ the comoving radial distance, and $\bmag$ and $\bevo$, respectively, the magnification and evolution bias. It is worth emphasising that \(\alpha\) is sample-dependent, due to the sample dependence of the two biases---just as the galaxy bias depends on the sample under consideration. For a luminosity-limited catalogue we have
\begin{equation}
\label{eq:bmag_bevo}
    \bmag=-\frac{\partial \ln{n(z;L>L_{\rm c})}}{\partial \ln{L_{\rm c}}} \;, \qquad
    \bevo=-\frac{\partial \ln{n(z;L>L_{\rm c})}}{\partial \ln{(1+z)}} \;,
\end{equation}
where $n(z;L>L_{\rm c})$ is the comoving (volumetric) number density of sources with a luminosity larger than a cut set at $L_{\rm c}$.

Moving to Fourier space, \cref{eq:delta_g_real_space} becomes
\begin{equation} \label{eq:delta_g_Fourier_space}
    \varDelta(\bm k)=
    % \left(b\,+f\,\mu^2+\ii\,\frac{\cH}{k}\,\alpha\,f\,\mu\right)
    \cZ^{(1)}(\bm k)
    \,\delta(\bm k)\;,
\end{equation}
where $\cZ^{(1)}(\bm k)=\cZ^{(1)}_{\rm N}(\bm k)+\cZ^{(1)}_{\rm GR}(\bm k)$ corresponds to the redshift-space kernel at first order in perturbation theory; its Newtonian (a.k.a.\ standard) and relativistic terms are
\begin{align}
    \cZ^{(1)}_{\rm N}(k,\mu)&=b\,+f\,\mu^2\;,\label{eq:Z1_N}\\
    \cZ^{(1)}_{\rm GR}(k,\mu)&=\ii\,\frac{\cH}{k}\,\alpha\,f\,\mu\;,\label{eq:Z1_GR}
\end{align}
with \reviews{$f\coloneqq-\de\ln D/\de\ln(1+z)$ the growth rate, given \(D\) the growth factor}, $\mu$ the cosine between the wavevector $\bm k$ and the line of sight, and \(k=|\bm k|\). Since the relativistic contribution (\ref{eq:Z1_GR}) to the kernel is proportional to $k^{-1}$, it mainly affects the largest scales, which are actually difficult to access due to cosmic variance. For this reason, the Doppler term has hitherto gone undetected, because of the lack of galaxy survey data on very large scales. 

Labelling tracers by $X$ and $Y$, the power spectrum tells us about their correlation in Fourier space and reads
\begin{multline} \label{eq:P_cross}
        P_{XY}(\bm k) = \Bigg\{\left(b_X\,+f\,\mu^2\right)\,\left(b_Y\,+f\,\mu^2\right)+\frac{\cH^2}{k^2}\,\alpha_X\,\alpha_Y\,f^2\,\mu^2\\
        + \ii\,\frac{\cH}{k}\,\left[\alpha_X\,\left(b_Y\,+f\,\mu^2\right) - \alpha_Y\,\left(b_X\,+f\,\mu^2\right) \right]\,f\,\mu \Bigg\} \, P(k) \;,
\end{multline}
with $P(k)$ the (linear) matter power spectrum and tracer-dependent quantities having subscripts $X$ and $Y$. Interestingly, we have $P_{XY}(\bm k)=P_{YX}(-\bm k)=P_{YX}^\ast(\bm k)\,$, where the asterisk denotes complex conjugation. If $X=Y$, i.e.\ in the case of auto-correlation, the imaginary term vanishes, thus the Doppler contribution becomes $\propto k^{-2}$; conversely, if $X \neq Y$ the relativistic effects leads to a dipole in the imaginary part, proportional to $k^{-1}$ \citep{2009JCAP...11..026M,2017JCAP...01..032G}. In other words, $b$ and $\alpha$ being different for the two tracers results in a cross-power spectrum which is more promising than the auto-correlation one and allows for a detection of the Doppler effect \citep{2023arXiv230604213B,2023arXiv230912400M}.

Being interested in the large scales, we rely on a simple Gaussian covariance matrix ${\sf\Gamma}$. Therefore, we define the power spectra data vector as ${\bm P}=(P_{XX},\,P_{XY},\,P_{YY})$ and consequently write the covariance associated with a measurement of such a multi-tracer galaxy power spectrum as
\begin{multline}\label{eq:variance1}
    {\sf\Gamma} = 
    \frac{2}{N_{\bm k}} \, \times \\ \left[
    \begin{array}{ccc}
    \tilde P_{XX}^2     & \tilde P_{XX}\,\tilde P_{XY}          & \tilde P_{XY}^2 \\
    \tilde P_{XX}\,\tilde P_{YX}   & \frac12\,\left(\tilde P_{XX}\,\tilde P_{YY}+ \tilde P_{XY}\,\tilde P_{YX}\right) & \tilde P_{XY}\,\tilde P_{YY} \\
    \tilde P_{YX}^2     & \tilde P_{YX}\,\tilde P_{YY} & \tilde P_{YY}^2
    \end{array}
    \right],
\end{multline}
where $\tilde P_{XY} = P_{XY} + N_{XY}$, $N_{XY}$ is the noise related to a measurement of $P_{XY}$ and $N_{\bm k}$ represents the number of independent modes available in the observed volume $V$. Technically, we compute this latter quantity by taking into account the width of $z-$, $\mu -$ and $k-$bins---which we dub $\Delta z$, $\Delta \mu$ and $\Delta k$, respectively---as well as considering $V(z,\Delta z)$ as the volume of a spherical shell rescaled by the observed area of the sky $f_{\rm sky}$, so that 
\begin{align}
    N_{\bm k}&=\frac{V(z,\Delta z)}{(2\,\pi)^{3}}\,2\,\pi\,k^2\,\Delta k\,\Delta\mu\;, \quad {\rm with} \nonumber \\
    V(z,\Delta z)&=\frac{4\,\pi f_{\rm sky}}3\left[r^3\!\left(z+\frac{\Delta z}2\right)-r^3\!\left(z-\frac{\Delta z}2\right)\right].
\end{align}
    
We stress the fact that we find \cref{eq:variance1} to be the correct and most general expression for a multi-tracer power spectrum in this form. When we work with imaginary contributions to the cross-correlation, we indeed need to deal with the complex nature of the off-diagonal terms and the above-mentioned relation $P_{XY}=P_{YX}^\ast$. As may be easily noticed, $\sf\Gamma$ is thus consistently Hermitian and non-negative, as it must always be for a covariance matrix. Also, our \cref{eq:variance1} coherently recovers both Eq.\ (2.5) of \citet{2024JCAP...02..043B} and Eq.\ (28) of \citet{2023EPJC...83..320J} in the case of a cross-power spectrum $P_{XY}$ being fully real (see also \citet{2024JCAP...03..034K} for a more in-depth analysis on multi-tracer covariance). 

\Cref{eq:delta_g_real_space} considers, on top of the Kaiser RSD correction, only the dominant term due to GR. However, other effects are expected to play a role \citep{2022JCAP...01..061C,2023PhRvL.131k1201F}. All those not included contributions, especially lensing and wide-angle effects \citet{2018MNRAS.476.4403C,2023JCAP...04..067P}, may be added in future analyses. We neglect these terms now for simplicity, because they cannot be directly added in a full power spectrum study. In addition to that, the very scales at which the Doppler contribution becomes dominant are also of a concern for constraining Primordial Non-Gaussianity, which in fact show a similar $k$-dependence and therefore needs to be distinguished from the relativistic effect \citep{2011PhRvD..83l3514N,2022A&A...662A..93E,2017PhRvD..96l3535A,2021JCAP...12..004V,2021JCAP...11..010V,2015MNRAS.451L..80C,2018PhRvD..97b3537L}. Nonetheless, since our main purpose is to understand whether we will be able to measure the Doppler effect within only one galaxy population, still through a multi-tracer approach, we shall assume the validity of \cref{eq:delta_g_real_space}, despite its non-completeness. 

%----------------------------------------------------------------------------------------------%
\section{Data sets} \label{sec:data}
\subsection{Luminosity cut technique}
Galaxy surveys typically provide us with flux/magnitude limited catalogues of redshift and angular position of sources. Such catalogues are essentially sets of galaxies, within given sky patches, that are observed with a flux density higher than the detector threshold. However, as was first proposed in \citet{2014PhRvD..89h3535B} (see also \cite{2016JCAP...08..021B,2017JCAP...01..032G} and the recent \cite{2023arXiv230604213B}), nothing forbids us to split such a set into two sub-samples, according to luminosity, and then treat the two sub-selections as if they were two independent populations.
Labelling with the subscript ``c" the critical flux density (or magnitude)---i.e.\  the detector threshold---and with ``s" the fixed division between the two sub-samples, we define a faint sample as the selection of all the sources whose observed brightness is within $[F_{\rm c},\,F_{\rm s})$---or analogously $\in (m_{\rm s},\,m_{\rm c}]$---and a bright sample which includes all the galaxies brighter than $F_{\rm s}$ (or $m_{\rm s}$). In line with the notation in \citet{2023arXiv230912400M}, we have $n_{\rm T}=n_{\rm F}+n_{\rm B}$, where $n_X$ is the galaxy number density for the faint ($X={\rm F}$), bright ($X={\rm B}$), or the total---namely, the whole catalogue---($X={\rm T}$) sample.

This luminosity cut immediately leads to the need to model the linear, magnification and evolution biases for the sub-sample, as they are different from those of the total population now. Specifically, we must describe the biases of the faint population ($b_{\rm F}$, $\bmag_{\rm F}$ and $\bevo_{\rm F}$) properly, considering that it is defined by both a lower and an upper flux limit. Conversely, the bright selection does not require any peculiar treatment because it can be thought of as a total catalogue with a more conservative luminosity threshold, which can be described by the usual recipes present in the literature. 
Therefore, we write the linear bias of the faint population as
\begin{equation}
    b_{\rm F}=\frac{n_{\rm T}\,b_{\rm T}-n_{\rm B}\,b_{\rm B}}{n_{\rm F}}\;,
\end{equation}
following \citet{2014MNRAS.442.2511F}. Given the number densities $n_{\rm T}$, $n_{\rm F}$ and $n_{\rm B}$, $b_{\rm F}$ only depends on the cumulative biases of the total and bright samples ($b_{\rm T}$ and $b_{\rm B}$) that in turn depends on $F_{\rm c}$ (or $m_{\rm c}$) and $F_{\rm s}$ (or $m_{\rm s}$), respectively.
Concerning the magnification bias $\bmag_{\rm F}$, it has to be derived by calculating the variation of the number of faint galaxies due to luminosity perturbation at both brightness limits \citep{2023arXiv230604213B}. This results in a redefinition of the bias in a way that takes into account also the logarithmic derivative of the galaxy number density with respect to the upper luminosity cut: 
\begin{equation} \label{eq:Q_F}
    \bmag_{\rm F} =\frac{n_{\rm T}}{n_{\rm T}-n_{\rm B}}\,\bmag_{\rm T} - \frac{n_{\rm B}}{n_{\rm T}-n_{\rm B}}\,\bmag_{\rm B} \;.
\end{equation}
On the other hand, since a redshift partial derivative is involved in the computation of the evolution bias, it is straightforward to rephrase it in the case of a faint sample as
\begin{equation} \label{eq:bevo_F}
    \bevo_{\rm F} =-\frac{\partial \ln{(n_{\rm T} - n_{\rm B})}}{\partial \ln{(1+z)}}\;.
\end{equation}

\subsection{Galaxy populations}
The relativistic Doppler contribution is sample-dependent, therefore the estimation of its detection probability cannot help but focus on specific targets. In this work, we consider the low redshift probe of a Bright Galaxy Sample and a high-$z$ H$\alpha$ emission line galaxy population. 

We take as the first case study the population of the bright near galaxies and model it after the Bright Galaxy Sample (BGS) of the Dark Energy Spectroscopic Instrument \citep{2023arXiv230606309S,2023AJ....165..253H}. The DESI BGS will target more than $10$ million galaxies at $z<0.6$ within a sky area of about $14\,000 \, \rm deg^2$ \citep[see also][]{2016arXiv161100036D,2024AJ....167...62D,2024arXiv240517208K}. Following the procedure illustrated in Appendix A.2 of \citet{2023arXiv230912400M}, we choose to model this galaxy population according to \citet{2023arXiv231208792S} and fix the critical (maximum) $r$-band magnitude for our analysis to $m_{\rm c}=20.175$. Briefly, we adopt the fit parameters given in \citet{2023arXiv231208792S} to build an analytical, luminosity-dependent Halo Occupation Distribution from which we consistently recover the galaxy number density and the values of the linear, magnification and evolution biases. In doing so, we assume the number of central galaxies within a dark matter halo to be modelled as an error function depending upon the halo mass, whose shape is given by the absolute magnitude of the galaxies themselves \citep{2005ApJ...633..791Z}. It is worth stressing that having a model which is able to describe the number of detected objects as a function of magnitude is crucial to our purpose: since we aim to work with a luminosity cut we indeed need to estimate the densities and the biases of the two magnitude-limited sub-samples. We work, in line with what is usually done in the case of the BGS, in a $ r$-magnitude-based framework, and use the K correction as in \citet{2021JCAP...04..055J}.

Further, we focus on a high redshift emission line galaxy population which mimics the H$\alpha$ target sample of the \textit{Euclid} satellite \citep{2011arXiv1110.3193L,2013LRR....16....6A,2018LRR....21....2A,2024arXiv240513491E}. \textit{Euclid} has recently started its scientific survey mapping up to $30$ million H$\alpha$ galaxies in $0.9 \lesssim z \lesssim 1.8$ over almost $15\,000 \, \rm deg^2$. We model such a population as the updated luminosity function given by \citet{2020A&A...642A.191E}, which is the \textit{Model 3} in \citet{2016A&A...590A...3P} within the reduced redshift range $z\in [0.9,\,1.8]$. We compute the evolution and magnification biases as was done in \citet{2021JCAP...12..009M} and consider a reference flux cut of $F_{\rm s}=2.0\fluxum$. On top of that, we parametrise the linear bias by means of the phenomenological formula presented in \citet{2020MNRAS.493..747P}, which can be employed in describing \textit{Euclid}-like H$\alpha$ targets.
For the sake of completeness, we point out that there is another luminosity function model, dubbed \textit{Model 1}, widely present in the literature \citep[see e.g.\ ][]{2020JCAP...03..065M,2021JCAP...12..009M,2023arXiv230912400M}. However, we shall not focus on that because its analytical formulation features a discontinuity at $z=1.3$ in $\bevo$ which may lead to unphysical outcomes. 

\begin{figure}
    \centering
    \includegraphics[width=\columnwidth]{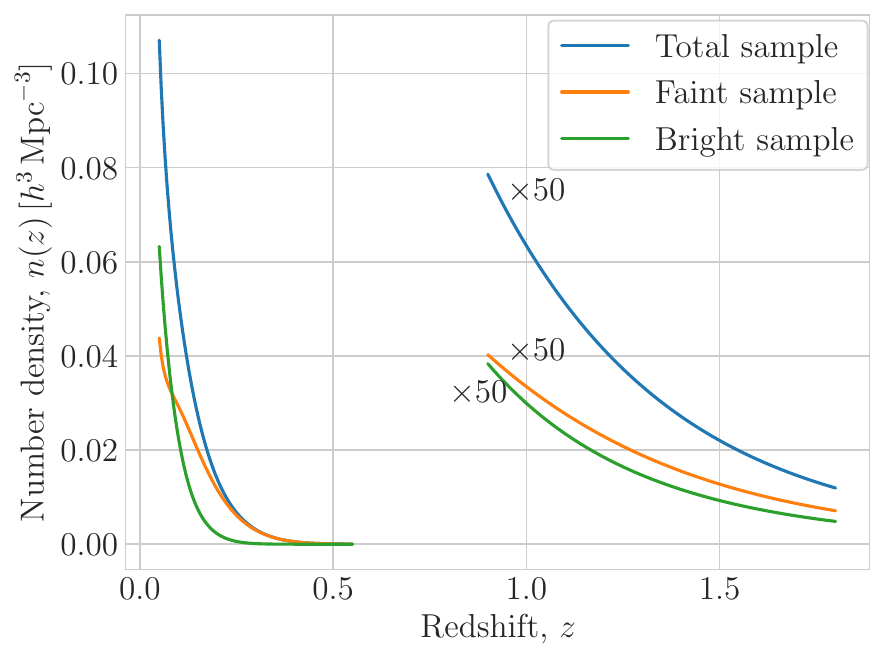}
    \caption{Galaxy number densities for the BGS (low-$z$) and H$\alpha$ (high-$z$) targets. Due to visualisation choice, curves for H$\alpha$ are multiplied by a factor of $50$. Critical and splitting magnitudes/fluxes are: $m_{\rm c}=20.175$ and $m_{\rm s}=19.0$ for the BGS; $F_{\rm c}=2.0\fluxum$ and $F_{\rm s}=2.8\fluxum$ for the H$\alpha$ population. }
    \label{fig:nz}
\end{figure}
\Cref{fig:nz} shows the galaxy number density for both galaxy populations after a flux cut between a faint and a bright sub-sample is applied. Within the redshift range $z\in[0.05,\,0.55]$ we display the BGS number densities referred to a magnitude split of $m_{\rm s}=19$; whereas in $z\in[0.9,\,1.8]$ H$\alpha$ curves, with $F_{\rm s}=2.8\fluxum$, are shown. A factor of $50$ amplifies the latter for a visualisation preference. Further, we leave in \ref{ap:Modelling} the fitting function that can be used to describe number densities and biases with the luminosity cut applied here. The choice of the division between the two sub-populations may seem arbitrary at this point, but it shall be clarified in \cref{sec:results}.

%----------------------------------------------------------------------------------------------%
\section{Information analysis} \label{sec:analysis}
In order to quantify the presence of a relativistic Doppler signal in the data, we carry out an information matrix analysis in several parameter spaces, which we shall discuss in the following section.
We start defining the auto- and cross-power spectra data vector as the vector that includes the three correlations, i.e.\ 
\begin{equation} \label{eq:datavector}
    {\bm P}(\bar z_i,\mu_m,k_n) = 
    \left[\begin{array}{c}
        P_{XX}(\bar z_i,\mu_m,k_n) \\
        P_{XY}(\bar z_i,\mu_m,k_n) \\
        P_{YY}(\bar z_i,\mu_m,k_n) \\
    \end{array}\right]
    \,.
\end{equation}
Then, from a purely theoretical perspective, given a parameter set $\bm \theta=\{\theta_1\cdots\theta_n\}$, which the power spectra depend upon, with $\{\theta_{\alpha}\}$ the $n$ so far unspecified parameters, we write the information matrix in a general way. It reads
\begin{multline}\label{eq:Fisher}
    I_{\alpha\beta}(\bar z_i)=\\
    \sum_{m,\,n}
    \frac{\pd {\bm P(\bar z_i,\mu_m,k_n)}^{\sf H}}{\pd \theta_{(\alpha}}\,{\sf\Gamma^{-1}}(\bar z_i,\mu_m,k_n)\,\frac{\pd {\bm P(\bar z_i,\mu_m,k_n)}}{\pd \theta_{\beta)}}\;,
\end{multline}
where the sum runs over all the configurations $(\mu_m,\,k_n)$ and parenthesis around indexes indicate the non-trivial symmetrisation which ensures ${\sf I}=\{I_{\alpha\beta}\}$ to be itself real. Indeed, it is worth noting that such symmetrisation is always required, as $\sf I$ is defined as the Hessian matrix of a Gaussian likelihood, but usually not made explicit when working with real power spectra.
In each redshift bin, the marginal errors on the parameters $\{\theta_\alpha\}$ are $\sigma_{\theta_\alpha}=\sqrt{({\sf I}^{-1})_{\alpha\alpha}(\bar z_i)}$. These values provide us with an estimation of the uncertainty of a measurement of ${\bm \theta}(\bar z_i)$ due to statistical noise and parameter degeneracies.
On the other hand, if we look at the entire $z$-range we obtain the cumulative marginal errors by taking the diagonal elements of the inverse of the total information matrix ${\sf I}=\sum_i{\sf I}(\bar z_i)$. 

Throughout this paper, we assume a standard $\Lambda \rm CDM$ \citet{2020A&A...641A...6P} cosmology in presenting our forecasts. Furthermore, we pick $f_{\rm sky}=0.36$, corresponding to a survey observing the majority of the extra-galactic sky. Such an assumption is common in literature and enables us to describe a \textit{Euclid}-like and a DESI-like survey at once, even though the nominal sky coverage is $14000 \, \rm deg^2$ ($f_{\rm sky}=0.339$) for the \textit{Euclid} satellite \citep{2024arXiv240513491E,2024arXiv240513492E} and for DESI \citep{2024AJ....167...62D}\footnote{Nonetheless, the assumption $f_{\rm sky}=0.36$ should not be seen as a purely optimistic hypothesis, because we expect both survey to be likely able to operate longer than the required time.}. This value is required for the computation of the probed volume and hence limits the largest measurable scale, which in turn enters our analysis as the smallest wavenumber considered, $k_{\rm min}$. Since we seek to detect an effect which is dominant on large scales, $k_{\rm min}$ plays a crucial role: we fix it directly from the volume by
\begin{equation}
    k_{\rm min}(\bar z_i,\Delta z)= \frac{2\,\pi}{\sqrt[3]{V(\bar z_i,\Delta z)}}\;,
\end{equation}
i.e.\ the wavenumber associated with the length scale of the $i$th redshift bin. On the other hand, the largest wavenumber is expected to not dramatically affect our results. We thus make the conservative choice to fix it to the scale at which the non-linear growth of structure becomes dominant at $z=0$, i.e.\ $k_{\rm max}=0.2\,h\,\rm Mpc^{-1}$, although there are recipes to describe it as a redshift dependent quantity and to push to mildly non-linear scales. 

\Cref{eq:Fisher} shows that we must define $z$-, $\mu$- and $k$-bins in order to estimate the presence of a relativistic signal in the data. Therefore, we use 10 $\mu$-bins in the range $\mu \in [-1,1]$ and 30 log-spaced bins in $k$ with $k_{\rm min} \le k \le k_{\rm max}$. As far as redshift bins are concerned, we divide the BGS and H$\alpha$ targets into 3 and 4 $z$-bins, respectively. In our study this results in a redshift bin width of $\Delta z \approx 0.17$ within the range $0.05\le z \le 0.55$ for the BGS \citep{2023arXiv230604213B} and $\Delta z = 0.225$ within $0.9 \le z \le 1.8$ for the H$\alpha$ (\textit{Model 3}-given) sample \citep{2020A&A...642A.191E}. We highlight that the width of the redshift bins constrains the probed volume and thus the $k_{\rm min}$ value. Such an aspect is reflected in the choice of considering thicker $\Delta z$ than the usual $\Delta z \approx 0.1$.

%----------------------------------------------------------------------------------------------%
\section{Results} \label{sec:results}
In this Section, we present the results of our information matrix analysis, obtained by exploring three different sets of parameters. We focus on the significance of the relativistic effects within the data, as a unique signal which we forecast across the whole redshift range. Afterwards, we investigate the possibility of disentangling the amplitude of the Doppler effect $\alpha_X(z)$ for both the faint and the bright sub-sample, individually. 

\subsection{Relativistic Doppler detection} \label{sec:resuls_1}
Firstly, we estimate the probability of detecting the relativistic Doppler effect by considering three dummy variables---$A_{\rm N}$, $A_{\rm K}$ and $A_{\rm D}$---as our parameter set $\bm{\theta}(\bar z_i)$. Those variables, whose fiducial values are all fixed to unity, act as amplitudes of the three leading contributions to the galaxy power spectrum: the real-space clustering signal, usually referred to as Newtonian ($A_{\rm N}$); the linear Kaiser Redshift Space Distortion term ($A_{\rm K}$) and the Doppler ($A_{\rm D}$). We then rephrase \cref{eq:Z1_N,eq:Z1_GR} to
\begin{equation}\label{eq:delta_g_with_ampl}
    \begin{split}
    \cZ^{(1)}_{X,\,\rm N}(k,\mu)&=A_{\rm N}\,b_X\,+A_{\rm K}\,f\,\mu^2\;,\\
    \cZ^{(1)}_{X,\,\rm GR}(k,\mu)&=\ii\,A_{\rm D}\,\frac{\cH}{k}\,\alpha_X\,f\,\mu\;,        
    \end{split}
\end{equation} 
and obtain the expression of the power spectra with the amplitudes. Thus, we set up our information matrix approach with the parameter set 
\begin{equation}
    \theta_\alpha=\left\{A_{\rm N},\, A_{\rm K},\, A_{\rm D}\right\}
\end{equation}
to compute the marginal error, cumulative on the entire redshift range, associated with a measurement of the Doppler amplitude. 
Since we are interested in finding a tailored sample to detect the relativistic effect, we focus this section on the dependence of $\sigma_{A_{\rm D}}$ upon the splitting flux/magnitude. For this reason, we decide to fold together all the $z$-bins, by considering the cumulative marginal errors instead of the differential ones, and study whether the curve $\sigma_{A_{\rm D}}(F_{\rm s})$/$\sigma_{A_{\rm D}}(m_{\rm s})$ shows a minimum. This would mean that there exists a sort of optimal division between the faint and bright sub-samples.

Secondly, we extend our analysis by including an additional set of nuisance parameters, on top of the three fictitious amplitudes. Specifically, we account for possible departures from Poissonian errors and add to $\bm \theta$ a number of $3\times(\rm{number \; of \;}z\rm{-bins})$ noise parameters, one per redshift bin and per power spectrum type ($P_{FF}$, $P_{FB}$ and $P_{BB}$) considered \citep[see e.g.\ ][]{2002MNRAS.333..730C,2010PhRvD..82d3515H,2017MNRAS.470.2566P,2020MNRAS.495..932G,2024A&A...683A.253E}. Therefore we take into account 
\begin{multline}
    \theta_\alpha^{\rm (w/ \; noise)}= \\ 
    \left\{A_{\rm N},\, A_{\rm K},\, A_{\rm D}, 
    \, N_{FF}^{(i)},\,\ldots,\, N_{FB}^{(i)},\,\ldots,\, N_{BB}^{(i)},\,\ldots,\right\}\;,
\end{multline}
where the fiducial value of \review{$N_{XY}^{i}=0$} and the superscript $(i)$, running over all the $z$-bins, means that we have an independent additional parameter for each bin. In doing so, we substitute 
$ P_{XY}(\bar z_i,\mu_m,k_n)\; \rightarrow  P_{XY}(\bar z_i,\mu_m,k_n) + N_{XY}^{(i)}$ back in \cref{eq:datavector} and then build an information matrix whose dimension is $[3\times(1+\rm{number \; of \;}z\rm{-bins})]^2$.

\begin{figure}
    \centering
    \includegraphics[width=\columnwidth]{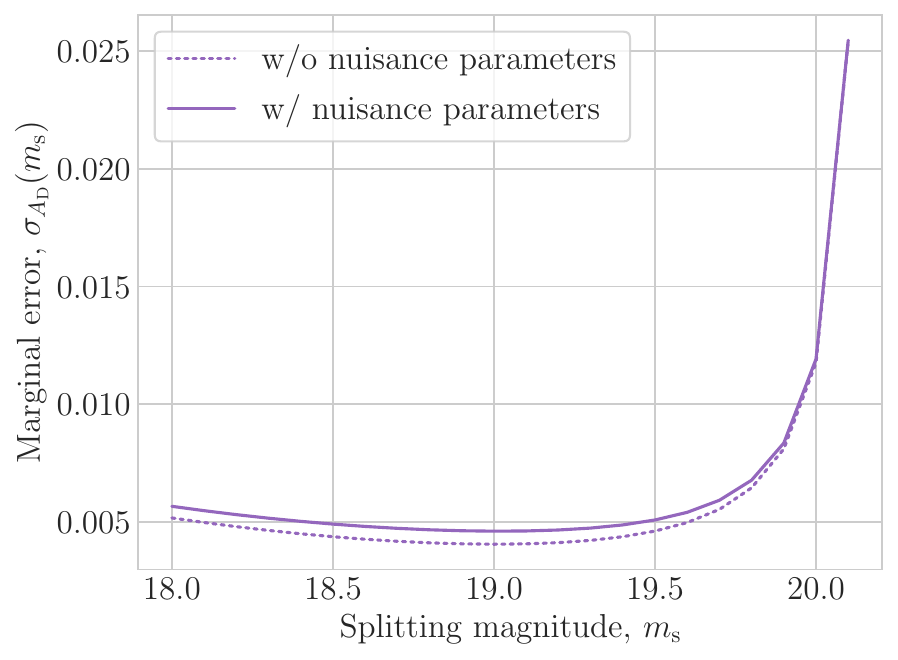}
    \caption{Cumulative marginal error associated with a measurement of the Doppler amplitude $A_{\rm D}$, within the entire redshift range, as a function of the splitting magnitude for the BGS. Solid(dashed) line refers to the analysis which includes(does not include) nuisance parameters, on top of the amplitudes $A_{\rm N}$, $A_{\rm K}$ and $A_{\rm D}$. Magnitude limit: $m_{\rm c}=20.175$.}
    \label{fig:fisherAD_BGS}
\end{figure}
BGS results are depicted in \cref{fig:fisherAD_BGS}, where we plot the cumulative marginal error associated with a measurement of the relativistic Doppler $A_{\rm D}$, as a function of the splitting magnitude. Being the fiducial value $A_{\rm D}=1$, we notice that the multi-tracer technique allows us to reach a high-significance detection, with $\sigma_{A_{\rm D}} \ll 1$. Also, the tiny discrepancy between the (solid) curve that includes the nuisance parameters and that (dashed line) which does not, demonstrate the robustness of our analysis and allows us to argue about the behaviour of the BGS sample itself. Finding $\sigma_{A_{\rm D}}(m_{\rm s})$ has a minimum at roughly $m_{\rm s}=19.0$, tells us it is possible to enhance the Doppler contribution within BGS data with the luminosity cut strategy. However, our purely analytical calculations might be non-accurate when $m_{\rm s}\lesssim 18.5$ because of the number density of the bright subsample $n_{\rm F}\, \rightarrow \,0$ in the highest-$z$ bin. For this reason, the accuracy for $\sigma_{A_{\rm D}}(m_{\rm s}\lesssim 18.5)$ is probably optimistic. 
\begin{figure}
    \centering
    \includegraphics[width=\columnwidth]{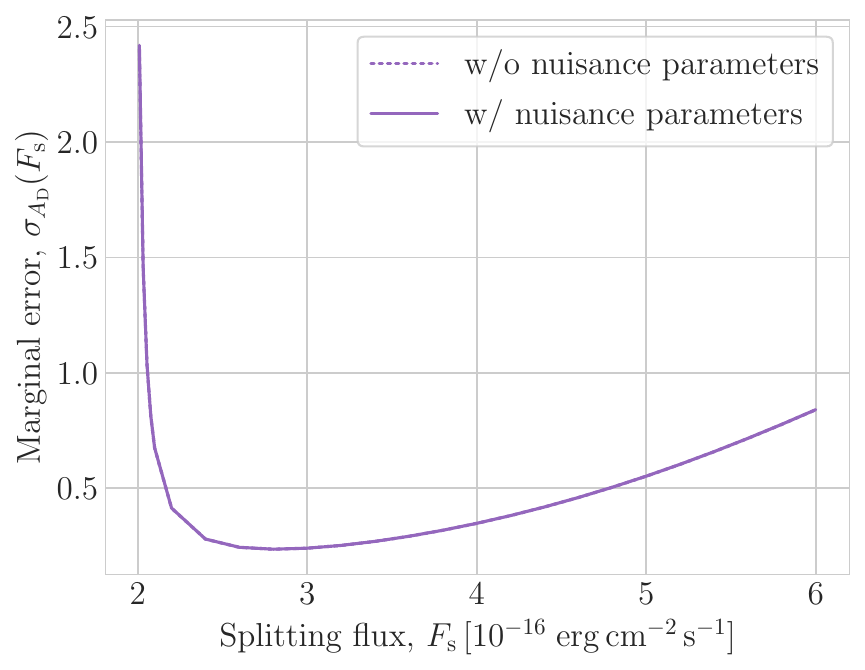}
    \caption{Same as \cref{fig:fisherAD_BGS} but in the case of the \textit{Euclid}-like H$\alpha$ sample: marginal error on $A_{\rm D}$ varying the splitting flux $F_{\rm s}$. Since a flux-based description is used, the faint end is on the left-hand side now (where the critical flux is $F_{\rm c}=2.0\fluxum$).}
    \label{fig:fisherAD_Euclid}
\end{figure}
Analogously, \cref{fig:fisherAD_Euclid} shows the $\sigma_{A_{\rm D}}(F_{\rm s})$ forecasts for the H$\alpha$ galaxy population. Again, results for both with- and without-noise cases are displayed; nevertheless, they almost overlap in our graph, being their relative difference $|\sigma_{A_{\rm D}}(F_{\rm s})^{(\rm w/o\;noise)}-\sigma_{A_{\rm D}}(F_{\rm s})^{(\rm w/\; noise)}|/\sigma_{A_{\rm D}}(F_{\rm s})^{(\rm w/\; noise)} < 1\%$. As expected, we find the marginal error on $A_{\rm D}$ marginalised over the nuisance parameters to be bigger than $\sigma_{A_{\rm D}}(F_{\rm s})^{(\rm w/o\;noise)}$. Both curves reach the $3\,\sigma$ detection level ($\sigma_{A_{\rm D}}\simeq0.3$) around $F_{\rm s}=2.8\fluxum$. H$\alpha$ galaxies allow for a detection of the relativistic effect with a lower significance than that given by the BGS due to the higher impact shown by the Doppler term at low redshift \citep{2020JCAP...07..048B}.

Moreover, if we look at the amplitudes of the clustering and Kaiser terms, namely $A_{\rm N}$ and $A_{\rm K}$, we notice that they are almost unaffected by the splitting flux/magnitude value. As a consequence, within the entire $F_{\rm s}$ (or $m_{\rm s}$) interval tested, they present roughly constant marginal errors, with discrepancies with respect to the average values well below $10\%$ (see \cref{tab:sigmaANAK} for details). The detection of the dominant terms of the power spectrum could indeed be affected by the noise alone, but it appears to be given by the number density of the total sample in a multi-tracer analysis, in which we merge information from both faint and bright sub-samples. 
\begin{table}
    \centering
    \begin{tabular}{rcc}
         \hline
         &                                          BGS & H$\alpha$\\
         \hline
         \hline
       $\sigma_{A_{\rm N}}^{\rm (w/o \; noise)}$  & $1.6\times 10^{-3}\pm 3.4\%$ & $7\times 10^{-4}\pm 0.2\%$\\
       $\sigma_{A_{\rm N}}^{\rm (w/ \; noise)}$  & $4\times 10^{-3}\pm 1.9\%$ & $6\times 10^{-4}\pm 0.2\%$\\
       $\sigma_{A_{\rm K}}^{\rm (w/o \; noise)}$  & $9\times 10^{-3}\pm 5.3\%$ & $3\times 10^{-3}\pm 0.3\%$\\
       $\sigma_{A_{\rm K}}^{\rm (w/ \; noise)}$  & $1\times 10^{-2}\pm 8.5\%$ & $1.4\times 10^{-3}\pm 0.4\%$\\
         \hline
    \end{tabular}
    \caption{Cumulative marginal errors on the Newtonian ($\sigma_{A_{\rm N}}$) and Kaiser ($\sigma_{A_{\rm K}}$) contributions for both BGS and H$\alpha$ targets. Superscripts refer to the presence/absence of noise parameters in the information matrix; whereas the reported values are expressed in the format $({\rm average \; value})\pm({\rm percentage \; variation})$, with the mean and variation being computed over the whole splitting flux/magnitude interval. Note that the uncertainty on $A_{\rm N}$ and $A_{\rm K}$ are hardly affected by the luminosity cut.}
    \label{tab:sigmaANAK}
\end{table}

Finally, it is worth noting that the results we present in this paper definitely look better than those obtained by focusing only on the faint-bright cross-correlation power spectrum, as can be appreciated if one compares \cref{fig:fisherAD_BGS,fig:fisherAD_Euclid} with Fig.\ 5 of \citet{2023arXiv230912400M}. That is due to the capability of the multi-tracer approach of putting all the information coming from both auto- and cross-power spectra together to get tighter constraints. This happens because, unlike the only cross-correlation case, a multi-tracer analysis is able to somehow overcome the cosmic variance limit, that always affects single spectra measurements \citep{2015ApJ...812L..22F}. 

\begin{figure*} 
    \centering
    \includegraphics[width=\textwidth]{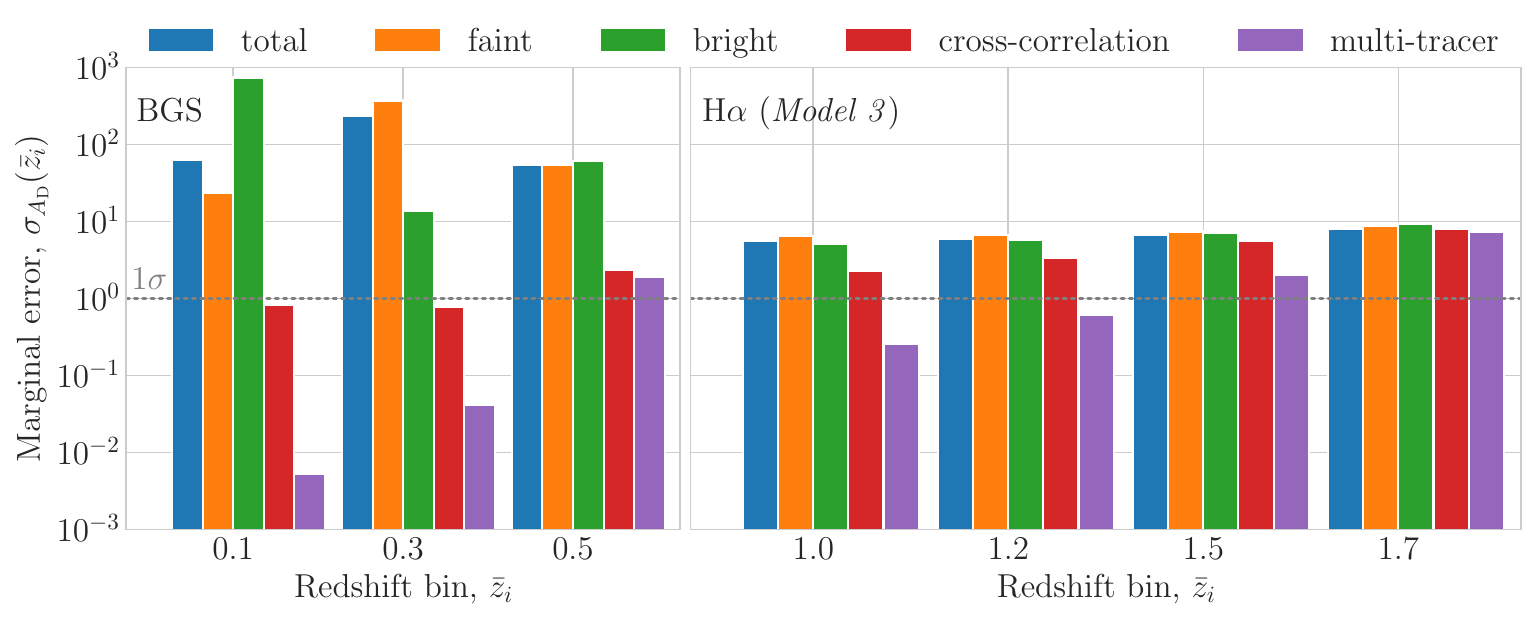}
    \caption{Differential marginal error associated with a measurement of $A_{\rm D}$ in each redshift bin, in the case of an optimal split between the two sub-samples (namely, $\{m_{\rm c},\,m_{\rm c}\}=\{20.175,\,19.0\}$ for the BGS and $\{F_{\rm c},\,F_{\rm s}\}=\{2.0,\,2.8\}\fluxum$ for H$\alpha$ galaxies). Blue, orange, green, and red candles respectively depict the cases of total, faint, and bright auto-power spectra and the faint-bright cross-power spectrum, whilst purple candles show the multi-tracer case.}
    \label{fig:sigmaAD_z}
\end{figure*}

To draw the reader's attention on how our cumulative marginal \reviews{errors turn into constraints} within each redshift bin, we show in \cref{fig:sigmaAD_z} the uncertainties in each redshift bin, $\sigma_{A_{\rm D}}(\bar z_i)$.\footnote{\Cref{fig:sigmaAD_z} can be seen as the information matrix complement of Fig.\ 2 in \citet{2023arXiv230912400M}, although there the detection significance of the Doppler contribution is depicted and the multi-\reviews{tracer} power spectrum is not yet considered.} We fix the luminosity cut to the best performing values, i.e.\ $m_{\rm s}=19.0$ and $F_{\rm s}=2.8\fluxum$ for the BGS and H$\alpha$ sample, respectively. Leaving aside the three auto-correlations (blue, orange, and green candles) that are unable to provide us with a detection of the relativistic contribution (see \ref{sec:auto-performance} for insights on the differences between those cases), we can appreciate how the marginal error improves for the cross-power spectrum (red candles) because of the different scaling of the Doppler \citep{2023arXiv230912400M,2023arXiv230604213B}. On top of that, the gain given by the multi-tracer measurements (purple candles) overcoming cosmic variance is remarkable. Thereby, \cref{fig:sigmaAD_z} demonstrates the importance of combining sample optimisation and multi-tracer analyses.

\subsection{Physical parameters}
Let us now move to estimate the uncertainty associated with a measurement of the following parameter set in each redshift bin (labelled by \(i\)),
\begin{equation}\label{eq:cosmopar}
    \reviews{\bm\theta}=\left\{\reviews{b\sigma_{8X,i},b\sigma_{8Y,i}},f\sigma_{8,i},\alpha_{X,i},\alpha_{Y,i}\right\}\;,
\end{equation}
where \reviews{\(b\sigma_{8X,i}\coloneqq b_X(\bar z_i)\,D(\bar z_i)\,\sigma_8\) and \(f\sigma_{8,i}\coloneqq f(\bar z_i)\,D(\bar z_i)\,\sigma_8\). This is a common parameterisation in template-fit analyses, and stems from the fact that, in a given redshift bin, both the bias and the growth rate are completely degenerate with the amplitude of the power spectrum in that bin; and, as such, they cannot be told apart in parameter estimation. As a consequence, we now express} the observable power spectra as
\begin{equation}\label{eq:P_cross_obs}
        P_{XY}(\bm k) = \reviews{\cZ^{(1)}_X(\bm k)\,\cZ^{(1)}_Y(-\bm k)\,}\frac{P(k)}{\reviews{D^2}\,\sigma_8^2}\;,
\end{equation}
with \reviews{the substitutions \(b_X\to b\sigma_{8X}\) and \(f\to f\sigma_8\) in the kernels of \cref{eq:Z1_N,eq:Z1_GR}. Finally, note that, as before, $X=Y$ yields} the auto-correlation case.

\reviews{Such a choice of} $\bm \theta$ addresses the problem from a more physical point of view. In this context we mimic a measurement of the two relativistic Doppler amplitudes in each $z$-bin. In other words, we take into account the difference between the biases of the faint and bright populations and no longer consider only one, general, amplitude. We thus fix the fiducial values of the parameters (\ref{eq:cosmopar}) to the theoretical predictions and estimate their uncertainties, focusing on the relativistic terms. 

We use once again the results shown above to select the best luminosity cut for both targets\reviews{,} i.e.\ $m_{\rm s}=19.0$ for BGS and $F_{\rm s}=2.8\fluxum$ for H$\alpha$ emitters.\footnote{As anticipated, these are the very divisions used in \cref{fig:nz}.}
\review{This choice allows us to study the relevance of the optimised Doppler signal in each sample. In fact, it is reasonable to assume that the best split found in \cref{sec:resuls_1} holds with the parameter set in \cref{eq:cosmopar} too. Since we are going to focus on $\alpha_{\rm F}$ and $\alpha_{\rm B}$---which together enclose the relativistic contribution---we expect their detection significance to peak in the same way $A_{\rm D}$ does.}

\begin{figure}
    \centering
    \includegraphics[width=\columnwidth]{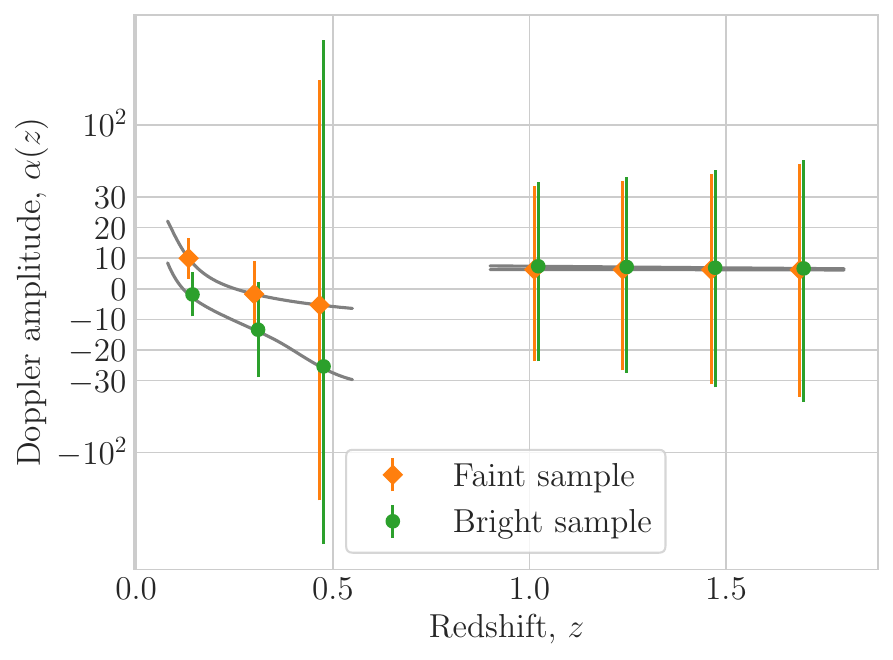}
    \caption{Doppler amplitude $\alpha(z)$, for the faint (orange) and bright (green) subsample in the best-split scenario. Results for the BGS are depicted on the left ($0.05\leq z \leq 0.55$) while H$\alpha$ curves lie on the right ($0.9\leq z \leq 1.8$). Coloured data points and error bars show our forecast about a measurement of the Doppler amplitude, i.e.\ they are given by $\alpha_{\rm F}\pm \sigma_{\alpha_{\rm F}}$ and $\alpha_{\rm B}\pm \sigma_{\alpha_{\rm B}}$ in each redshift bin; on the other hand, grey lines draw the theoretical predictions. Let us remark that, to facilitate visualisation, on the $y$-axis a `symlog' scale is adopted, linear for \reviews{$\alpha\in[-30,30]$} and mirrored logarithmic outside that range. }
    \label{fig:alpha_results}
\end{figure}
Illustratively, \cref{fig:alpha_results} shows our results of both BGS (low-$z$, left-hand side of the panel) and H$\alpha$ population (high-$z$, right-hand side). The Doppler amplitude $\alpha_X(z)$ is plotted against redshift using a symmetric logarithmic scale (linear in the range \reviews{$-30\le\alpha\le30$}). Grey curves represent the predictions for both sub-samples, whilst coloured data points (faint sample in orange, bright sample in green) are the measurement forecasts. Each data point is associated with an error that is given by the study of the information matrix, namely $\sigma_{\alpha_X}(\bar z_i)$. We find in general large error bars, meaning that it is much harder to constrain $\alpha_{\rm F}(z)$ and $\alpha_{\rm B}(z)$ in each redshift bin separately than assess the total presence of relativistic Doppler in the data. Moreover, we observe that the Doppler amplitude is well-constrained within the first two BGS redshift bins, whereas this is not the case with the third $z$-bin, due to the number density falling off at $z \gtrsim 0.35$. At high redshift, the difference between the faint and bright sub-samples is hardly appreciable, meaning that the bias corrections of \cref{eq:Q_F,eq:bevo_F} induced by the luminosity cut are more relevant for the BGS. This once again confirms the sample-dependent nature of the relativistic Doppler effect and, in turn, the need for a search for tailored galaxy samples in the efforts to detect this effect.

%----------------------------------------------------------------------------------------------%
\section{Discussion and conclusions}\label{sec:conclusions}
A detection of an effect due to GR coming from galaxy clustering measurements might further confirm the validity of Einstein's theory of gravity in the very cosmological regime where it has not been tested yet. Several corrections affect the relation between the (underlying) dark matter density contrast and that of galaxies \citep{2010PhRvD..82h3508Y,2011PhRvD..84f3505B,2011PhRvD..84d3516C}. Among them, the so-called relativistic Doppler term plays a dominant role with respect to the other contributions, especially at low redshift where lensing is subdominant \citep{2022JCAP...01..061C}. Its relevance is given by the amplitude parameter ($\alpha$), which behaves differently for different galaxy populations due to the presence of magnification and evolution biases. Moreover, the Doppler scaling $\propto k^{-1}$ in $\varDelta({\bm k})$ makes it important on large scales in the auto-correlation galaxy power spectrum and relevant even at intermediate scales in the cross-correlation case. The main issues to be addressed thus are: accessing the largest scales plagued by low statistical sampling due to the lack of Fourier modes, and finding a promising target. 

In the search for an optimal galaxy sample to achieve a detection of the relativistic Doppler term, we have split a galaxy population according to luminosity \citep{2014PhRvD..89h3535B}, and then theoretically estimated the significance of the relativistic signal in a multi-tracer power spectrum analysis \citep{2004MNRAS.347..645P}. To exploit multi-tracer benefits out of a single data set we have combined the auto-correlation of the two (faint and bright) sub-samples and their cross-power spectrum. We have presented forecasts regarding a low-$z$ DESI-like BGS and a high-$z$ \textit{Euclid}-like H$\alpha$ target. In doing so, we have coherently retrieved all the relevant quantities thanks to phenomenological relations calibrated on real data \citep{2020MNRAS.495.1340F,2023arXiv230912400M}.

Our information analysis approach has shown that we can tune the division between the faint and bright samples in order to maximise the impact of the relativistic effect. Considering the entire redshift range, i.e.\ the cumulative marginal error on the Doppler contribution $\sigma_{A_{\rm D}}$, we have found the following best selections:
\begin{itemize}
    \item $m_{\rm s}=19.0$ (with $m_{\rm s}=20.175$) for BGS;
    \item $F_{\rm s}=2.8\fluxum$ (with $F_{\rm c}=2.0\fluxum$) for H$\alpha$ emitters.
\end{itemize}
Interestingly, both cases roughly correspond to a $50\%$ faint-$50\%$ bright division. This outcome, regarding BGS, is not in agreement with the optimal $10\%$ faint-$90\%$ bright division obtained by \citet{2023arXiv230604213B}; however, we point out that such a difference might be given by the flattening of their number density at high $m_{\rm c}$ values, which we have not seen in our HOD study. 
The BGS has turned out to be more promising than the high-$z$ H$\alpha$ population, exhibiting a minimum $\sigma_{A_{\rm D}}\simeq0.005$, to be compared with the minimum H$\alpha$ value $\sigma_{A_{\rm D}}\simeq0.3$. This result may be due to both the larger number density of sources in the BGS and the $\alpha$'s specific features. 
Furthermore, we have checked the robustness of the forecasts presented by introducing various nuisance parameters within our information matrix and verifying that they do not have any significant impact on our findings. 
Also, we have used the most promising samples to make forecasts on the measurement of the Doppler amplitude in the two sub-samples. As expected on the basis of our previous results, $\alpha$ seems to be better constrained in the case of the BGS, particularly in the low-redshift bins, even though the error bars are such that we cannot forecast a clear joint detection of $\alpha_{\rm B}$ and $\alpha_{\rm F}$ yet.
To study the Doppler presence within the full galaxy power spectrum the computation of the corresponding covariance matrix has been required. The covariance treatment presented has been carried out for the first time taking into account the relativistic contribution being imaginary. Since $P_{XY}({\bm k})=P_{YX}^\ast({\bm k})$, a Hermitian covariance matrix indeed allows us to work with complex data vectors.

To conclude, thanks to the improved sensitivity and the enhanced volume probed, the current galaxy surveys \citep{2024arXiv240513491E,2024AJ....167...62D} are likely to provide us with a first detection of a relativistic signature on the large scales of the cosmic structures. As those effects are sample-dependent, accurate sample optimisation works will have a paramount relevance for the target selection (see e.g.\  \citet{2023arXiv230604213B} and \citet{2024arXiv240619908S}).
However, in order to obtain more accurate forecasts on the probability of detecting a GR-driven term, rather than focusing on the relevance of the relativistic Doppler itself, future studies shall include other contributions. In particular, it has been shown that the wide-angle effect should be considered as they are in fact not negligible with respect to the Doppler \citep{2023JCAP...04..067P,2018MNRAS.476.4403C,2023PhRvD.107h3528N,2024arXiv240606274J}. Additionally, Primordial Non-Gaussianity should be included \citep{2017PhRvD..96l3535A,2015MNRAS.451L..80C,2012PhRvD..85d1301B}, as well as the other local and integrated effects in the full relativistic correction to the galaxy number density in redshift space \citep{2023PhRvL.131k1201F,2022JCAP...01..061C,2020JCAP...07..048B}. Extensions of this study are going to focus on the analysis of the faint-bright multi-tracer power spectrum in harmonic space \citep{TesiMarco} and assess the reliability of the luminosity cut technique using simulated data.

%----------------------------------------------------------------------------------------------%
\section*{Acknowledgments}
The authors \review{are grateful to the anonymous reviewer for their encouraging review and for their comments, which helped to improve the quality of this paper.} They also warmly thank S.J.\ Rossiter for reading the first version of the manuscript.
They acknowledge support from the Italian Ministry of University and Research (\textsc{mur}), PRIN 2022 `EXSKALIBUR – Euclid-Cross-SKA: Likelihood Inference Building for Universe's Research', Grant No.\ 20222BBYB9, CUP C53D2300131 0006, from the Italian Ministry of Foreign Affairs and International
Cooperation (\textsc{maeci}), Grant No.\ ZA23GR03, and from the European Union -- Next Generation EU.

%----------------------------------------------------------------------------------------------%
\appendix \section{Fits} \label{ap:Modelling}
We report here the fitting functions for the number densities and the biases with the optimal luminosity cut, in order to aid the reproducibility of our results.

%Fit in BGS_biasiipynb
\subsection{BGS}\label{sec:BGS_fits}
The fitting formulae for the total, bright and faint samples with $m_{\rm c}=20.175$ and $m_{\rm s}=19.0$ are given by (all fitting formulae are for comoving number densities in units of \(h^3\,{\rm Mpc^{-3}}\)):
\begin{itemize}
    \item $n_{\rm T}(z)=0.211 $ \\ $+ \left(-2.77\,z+4.80\,z^2-13.6\,z^3\right)\,e^{-4.39\,z}$;
    \item $n_{\rm F}(z)=0.0606$ \\ $ + \left(-0.389\,z+0.797\,z^2-0.557\,z^3\right)\,e^{-0.150\,z}$;
    \item $n_{\rm B}(z)=0.151 + \left(-2.45\,z+5.52\,z^2-14.0\,z^3\right)\,e^{-4.72\,z}$;
    \item $b_{\rm T}(z)=1.2-0.7\,z+1.9\,z^2+9.5\,z^3$;
    \item $b_{\rm F}(z)=1.2-2.0\,z+6.2\,z^2+5.6\,z^3$;
    \item $b_{\rm B}(z)=1.2-2.4\,z+11.3\,z^2+22.5\,z^3$;
    \item $\bmag_{\rm T}(z)=0.2+4.2\,z+0.3\,z^2+19.3\,z^3$;
    \item $\bmag_{\rm F}(z)=0.3+17.7\,z^2+0.9\,z^3$;
    \item $\bmag_{\rm B}(z)=-0.2+15.2\,z-44.9\,z^2+140.7\,z^3$;
    \item $\bevo_{\rm T}(z)=-0.4-2.7\,z-3.5\,z^2-12.6\,z^3$;
    \item $\bevo_{\rm F}(z)=-0.3-0.9\,z-12.6\,z^2-2.4\,z^3$;
    \item $\bevo_{\rm B}(z)=0.4-21.3\,z+84.3\,z^2-177.6\,z^3$.
\end{itemize}

%Fit in CrossCorrelation-FaintBright_v5.ipynb
\subsection{H\(\alpha\) sample}\label{sec:ELG_fits} %CrossCorrelation-FaintBright_FisherMatrix_v5.ipynb 
On the other hand, we write in this section only the number densities, $\bmag$ and $\bevo$, being the function for the linear bias of the total sample already included in \citet{2020MNRAS.493..747P}. Therefore, the fitting formulae with $F_{\rm c}=2.0\fluxum$ and $F_{\rm s}=2.8\fluxum$ (again, all fitting formulae for comoving number densities are expressed in units of \(h^3\,{\rm Mpc^{-3}}\)) are:
\begin{itemize}
    \item $n_{\rm T}(z)=0.00767-0.01116\,z+0.00584\,z^2-0.00107\,z^3$;
    \item $n_{\rm F}(z)=0.00326-0.00427\,z+0.00202\,z^2-0.00034\,z^3$;
    \item $n_{\rm B}(z)=0.00441-0.00690\,z+0.00382\,z^2-0.00073\,z^3$;
    \item $\bmag_{\rm T}(z)=+0.59+2.01\,z-0.57\,z^2+0.04\,z^3$;
    \item $\bmag_{\rm F}(z)=-0.08+2.48\,z-0.62\,z^2+0.02\,z^3$;
    \item $\bmag_{\rm B}(z)=+0.77+2.58\,z-1.17\,z^2+0.19\,z^3$;
    \item $\bevo_{\rm T}(z)=-7.50+0.86\,z+0.95\,z^2-0.25\,z^3$;
    \item $\bevo_{\rm F}(z)=-3.43-4.66\,z+3.63\,z^2-0.71\,z^3$;
    \item $\bevo_{\rm B}(z)=-10.03+2.93\,z+0.59\,z^2-0.29\,z^3$.
\end{itemize}

\section{Insights on sub-sample auto-correlation performance} \label{sec:auto-performance}
It may appear counter-intuitive that a sub-sample can perform better than the total sample, when it comes to detect some effect, as in our case. In particular, in \cref{fig:sigmaAD_z}
we see that the faint sub-sample alone yields tighter constraints on \(A_{\rm D}\) in the first BGS redshift bin than not using the total sample. The same happens for the bright sample in the second bin. Here, we give an analytical demonstration of how this can happen in practice. 

Let us compare the information matrix of the total sample, \({\sf I}^{\rm T}\), with any of those of its sub-samples, \({\sf I}^J\), where \(J=\{{\rm F},{\rm B}\}\). For simplicity, let us also consider the two limiting cases of cosmic-variance domination (CVD) and shot-noise domination (SND); this will turn out handy because, as we shall see later, the dependence on the Fourier wavenumber \(k\) cancels out exactly. Focussing on \(A_{\rm D}\) alone, we have that \cref{eq:Fisher} becomes, after some simplifications,
\begin{multline}
    I^X_{A_{\rm D}A_{\rm D}}=2\,\sum_{m,n}N_{\bm k}\,\left(\frac\cH{k_n}\,\alpha_X\,f\,\mu_m\right)^4 \\
    \times\begin{cases}
        \left[\left(b_X+f\,\mu_m^2\right)^2+\left(\cH\,\alpha_X\,f\,\mu_m/k_n\right)^2\right]^{-2} & \text{(CVD)} \\
        P^2(k_n)\,n_X^2 & \text{(SND)}
    \end{cases}\;,
\end{multline}
where we have omitted redshift dependence for simplicity.

Now, a sub-sample performs better than the total sample if and only if \({\sf I}^J>{\sf I}^{\rm T}\). This is an inequality that can be solved analytically for \(\alpha_J\) at fixed \(k_n\) and \(\mu_m\), namely
\begin{equation}
    |\alpha_J|>
    \begin{cases}
        |\alpha_{\rm T}|\,\left(b_J+f\,\mu_m^2\right)/\left(b_{\rm T}+f\,\mu_m^2\right) & \text{(CVD)} \\
        |\alpha_{\rm T}|\,\sqrt{n_{\rm T}/n_J} & \text{(SND)}
    \end{cases}\;,\label{eq:alpha_conditions}
\end{equation}
where \(|\alpha_{\rm T}|\) emphasises the fact that \(\alpha_{\rm T}\) can also be negative, whilst \(b_X+f\,\mu^2\) and \(n_X\) are clearly strictly positive. Since \(k_n\) appears no longer in \cref{eq:alpha_conditions}, if the conditions are satisfied on some comoving scale, they are satisfied everywhere. Furthermore, in SND the condition does not depend anymore on \(\mu_m\) either. In our case (it can be checked by using the fitting formulae in \ref{sec:BGS_fits} and \ref{sec:ELG_fits}), we find that the conditions \cref{eq:alpha_conditions} for SND are satisfied in both the first and the second redshift bin:
\begin{enumerate}
    \item \reviews{\(|\alpha_{\rm F}|>7.5\), with \(\alpha_{\rm F}=10.8\)};
    \item \reviews{\(|\alpha_{\rm B}|>8.6\), with \(\alpha_{\rm B}=-13.4\)}.
\end{enumerate}

\begin{figure}
    \centering
    \includegraphics[width=\columnwidth]{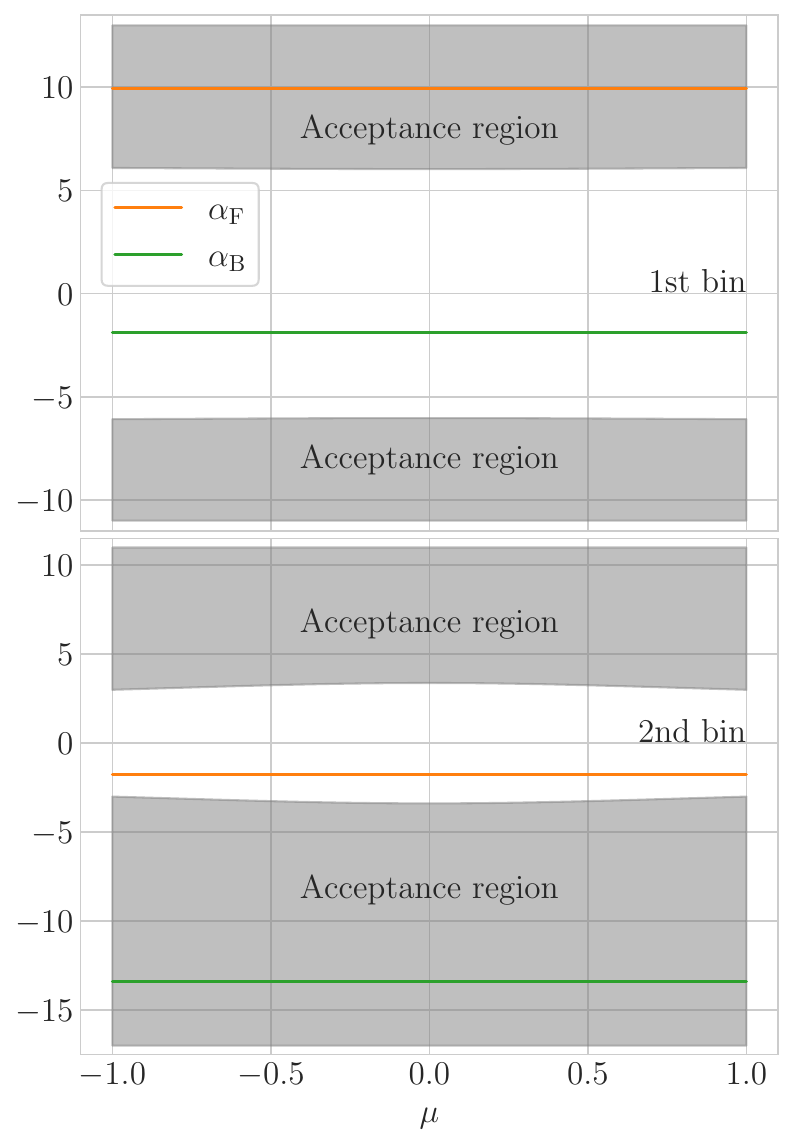}
    \caption{Acceptance regions (shaded grey areas) for the \({\sf I}^J>{\sf I}^{\rm T}\) conditions \cref{fig:CVD_condition} in CVD. The top and bottom panels respectively refer to the first and second redshift bin for the BGS, whilst orange and green lines mark the value of \(\alpha_J\). In accordance to what shown in \cref{fig:sigmaAD_z}, in the first bin the faint sample meets the aforementioned conditions, whereas the bright sample does not; the reverse occurs in the second redshift bin.}
    \label{fig:CVD_condition}
\end{figure}

In the case of CVD, there is still a datum-by-datum dependence upon the line of sight through the RSD term \(\propto\mu_m^2\). But, again, if we prove that the CVD inequality is satisfied for any value of \(\mu\), the condition will be always satisfied. The result of this exercise is presented in \cref{fig:CVD_condition}, as a function of \(\mu\). This demonstrates that it is indeed possible to obtain a better performance from a sub-sample, despite a lesser statistics, due to the impact the choice of the sub-sample has on the signal we are after.

%----------------------------------------------------------------------------------------------%

\bibliographystyle{apsrev4-1}
\bibliography{apssamp}% Produces the bibliography via BibTeX.

\end{document}